# Application of Graph Coloring to Biological Networks

Susan Khor

**Abstract**

We explore the application of graph coloring to biological networks, specifically protein-protein interaction (PPI) networks. First, we find that given similar conditions (i.e. number of nodes, number of links, degree distribution and clustering), fewer colors are needed to color disassortative (high degree nodes tend to connect to low degree nodes and vice versa) than assortative networks. Fewer colors create fewer independent sets which in turn imply higher concurrency potential for a network. Since PPI networks tend to be disassortative, we suggest that in addition to functional specificity and stability proposed previously by Maslov and Sneppen (Science 296, 2002), the disassortative nature of PPI networks may promote the ability of cells to perform multiple, crucial and functionally diverse tasks concurrently. Second, since graph coloring is closely related to the presence of cliques in a graph, the significance of node coloring information to the problem of identifying protein complexes, i.e. dense subgraphs in a PPI network, is investigated. We find that for PPI networks where 1% to 11% of nodes participate in at least one identified protein complex, such as *H. sapien* (DIP20070219, DIP20081014 and HPRD070609), DSATUR (a well-known complete graph coloring algorithm) node coloring information can improve the quality (homogeneity and separation) of initial candidate complexes. This finding may help to improve existing protein complex detection methods, and/or suggest new methods.

**Supplementary Material**

**SM-1 Supplementary Material for Section 2**

*Network formation*

Using a different random number seed each time, two networks with power-law distributed degree distributions are produced with the preferential attachment algorithm described in [2]. For both networks, all nodes belong to the same component, the number of nodes N = 1,000, and the number of links M = 4,960. Let these two networks form a set called D0. The relevant characteristics of these networks are given in Table SM-1.1 and Fig. SM-1.1.

**Table SM-1.1** Node degree summary statistics for the networks.

| Min | Max | Average | Std. dev. | Mod | Median |
|-----|-----|---------|-----------|-----|--------|
| 3   | 116 | 9.92    | 10.3947   | 5   | 7      |
| 5   | 102 | 9.92    | 9.2080    | 5   | 7      |

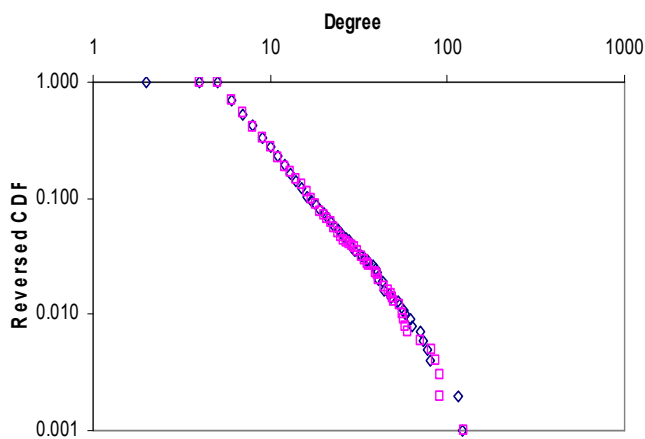

**Fig. SM-1.1** The reversed cumulative degree distributions of the test networks on a log-log scale.

In the first experiment (E1), assortative and disassortative versions of the networks in D0 are formed by rewiring randomly chosen pairs of links either to increase or to decrease degree-degree correlation per [20]. These networks have little to no clustering. In E1, the networks in D0 form the baseline or *null* model.

In the second experiment (E2), the node degree lists (which is a list of node degrees in node label order) of the networks in D0 are fed into the algorithm in [9] to produce networks with high clustering. Two networks are produced for each node degree list with a different random number seed each time. Let these four networks form a set called S0. In E2, the networks in S0 form the baseline or *null* model. Disassortative and assortative versions of the four networks in S0 are produced using the algorithm in Appendix A of [9] which essentially controls the links between the top 5% of high degree nodes. For E2, the link probability between the set of top 50 (5% × 1000) high degree nodes is set at 0.00 to create networks more disassortative than the null networks, and 0.25 and 0.75 to create networks more assortative than the null networks. Fig. SM-1.2 compares the clustering [19] and assortativity [12] characteristics of the E1 and E2 networks.

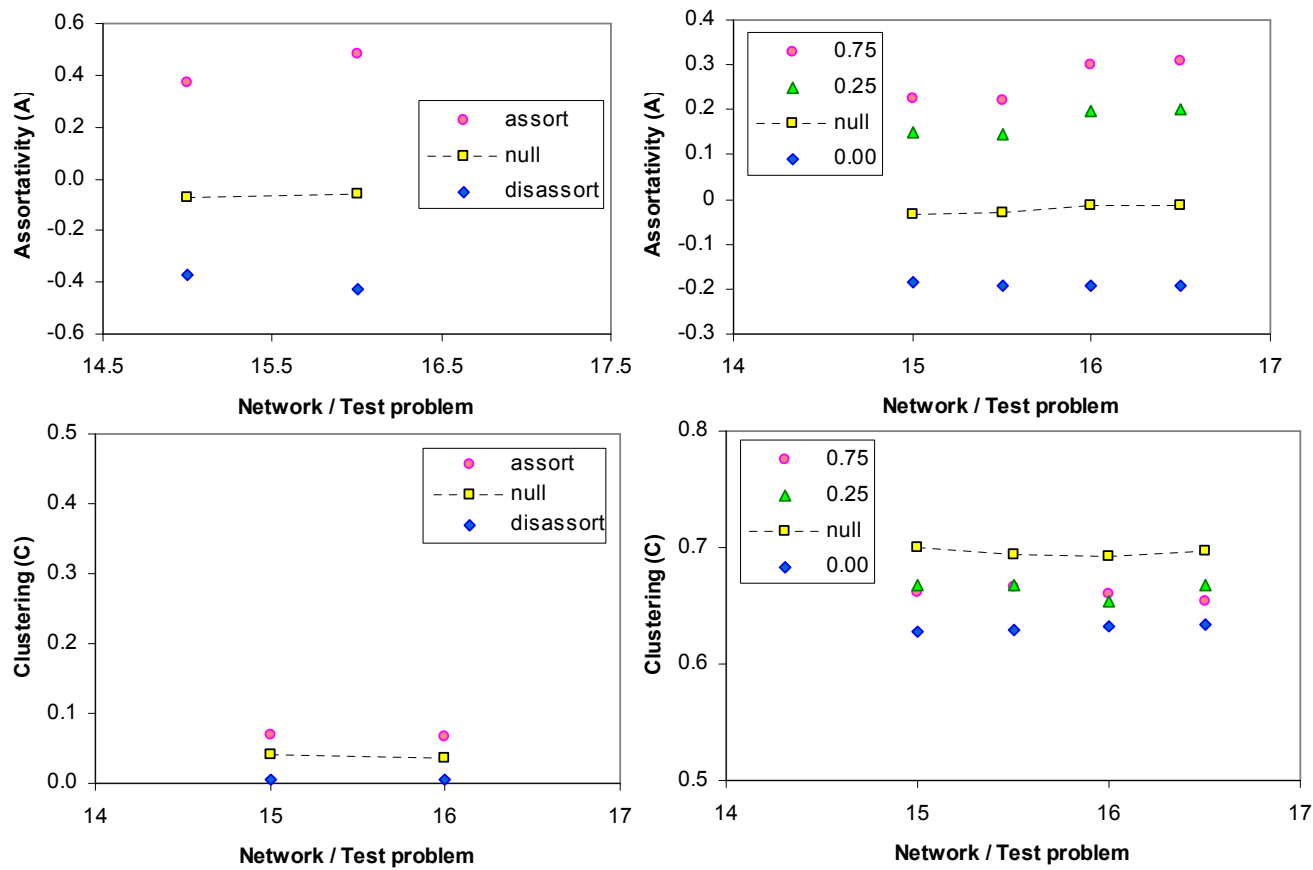

**Fig. SM-1.2** Topological characteristics of the two networks for E1 (left) and the four networks for E2 (right). The degree distributions of these networks are given in Fig. SM-1.1. The E2 networks labeled 15 and 15.5 (16 and 16.5) have the same degree distribution as the E1 network labeled 15 (16).

*Graph Coloring Algorithms*

The DSATUR (degree saturation) algorithm [3] begins by labeling a highest degree node with the lowest numbered color and proceeds to color one node at a time, giving preference to nodes of high saturation or of high degree if there is more than one node with the same amount of saturation, with the lowest numbered color without incurring a conflict. Saturation refers to the unique number of colors neighbouring an uncolored node. In our implementation, colors begin at 0 and increase by 1. We do not fix the number of colors $c$ for a network beforehand, but instead use DSATUR to find *c*. Thus, the *c* value found may or may not be the chromatic number of a network. DSATUR is run once per network.

The hill climbing (HC) algorithm repeatedly chooses a random number of nodes from a network to mutate, i.e. change to a randomly chosen color within a given palette, until either the network is properly colored or the maximum number of tries (max_evals) is reached. In the experiments, max_evals is set to 2 million. The number of nodes to mutate is controlled by the mutation rate (Pm), which in the experiments is set to 0.0625, permitting HC to mutate 1 to 62 (0.0625 × N) nodes at a time. In HC the current network is reproduced with some slight random variation via mutation and the better colored or fitter network of the parent-offspring pair is selected for reproduction in the next iteration while the less fit network is discarded. HC graph coloring is done by first using the number of colors required by DSATUR, and then as necessary, incrementing the number of colors until HC achieves a high (close to 100%) success rate, i.e. finds a proper coloring within max_evals on every run it does.

*Method*

DSATUR is run once per network and its results are averaged over network type, i.e. disassortative, null and assortative for E1, and 0.00, null, 0.25 and 0.75 for E2. Due to HC's stochastic nature, 10 independent runs (with a different random number seed each time) are made for each network, and results are averaged over all runs per network type. Unlike DSATUR, there is no inherent order in HC's color assignments, i.e. the highest degree node need not be labeled with color 0, and HC may produce different but proper *c*-coloring of a network. This difference between algorithms is considered when evaluating the results. Table SM-1.2 illustrates the result summarization process for Fig. 1.

Table SM-1.2

| Network | 15.0 | 16.0 | | Avg. | Network | 15.0 | 15.5 | 16.0 | 16.5 | Avg. |
|---|---|---|---|---|---|---|---|---|---|---|
| E1 | DSATUR Colors | | | | E2 | DSATUR Colors | | | | |
| Disassort | 6 | 6 | | 6 | 0.00 | 7 | 8 | 8 | 9 | 8.00 |
| Null | 7 | 7 | | 7 | Null | 9 | 10 | 11 | 11 | 10.25 |
| assort | 24 | 22 | | 23 | 0.25 | 13 | 12 | 13 | 12 | 12.50 |
| | | | | | 0.75 | 16 | 16 | 17 | 16 | 16.25 |

| Colors | 6 | 8 | 12 | 14 | 24 | 26 | Colors | 8 | 10 | 14 | 16 | 18 | 20 |
|---|---|---|---|---|---|---|---|---|---|---|---|---|---|
| E1 | RMHC Success Rate | | | | | | E2 | RMHC Success Rate | | | | | |
| Disassort | 20/20 | | | | | | 0.00 | 17/40 | 40/40 | | | | |
| Null | 0/20 | 0/20 | 16/20 | 20/20 | | | Null | | 0/40 | 22/40 | 37/40 | | |
| assort | | | | | 14/20 | 20/20 | 0.25 | | | | 32/40 | 39/40 | |
| | | | | | | | 0.75 | | | | | 36/40 | 39/40 |

*Results*

Fig. SM-1.3 examines the coloring of the top 50 high degree nodes. The DSATUR values are the average (avg) and one standard deviation (sd) of color values for the top 50 high degree nodes of each network. A low average combined with a small standard deviation indicates little variability in the coloring of the top 50 high degree nodes. This simple summary is not applicable to HC because unlike DSATUR, HC does not assign the lowest numbered color to nodes. Further, permutation of a proper coloring is also a proper coloring. Therefore, for HC, the one standard deviation value of color values for the top 50 high degree nodes of the 10 random runs is recorded, and the HC plots report the average of these standard deviations to indicate the color range of the top 50 high degree nodes. What is important is not the predominant color of the nodes of a network, but the number of or range of colors of the nodes, which tells us the number of independent sets and thus the groups of tasks that may execute concurrently.

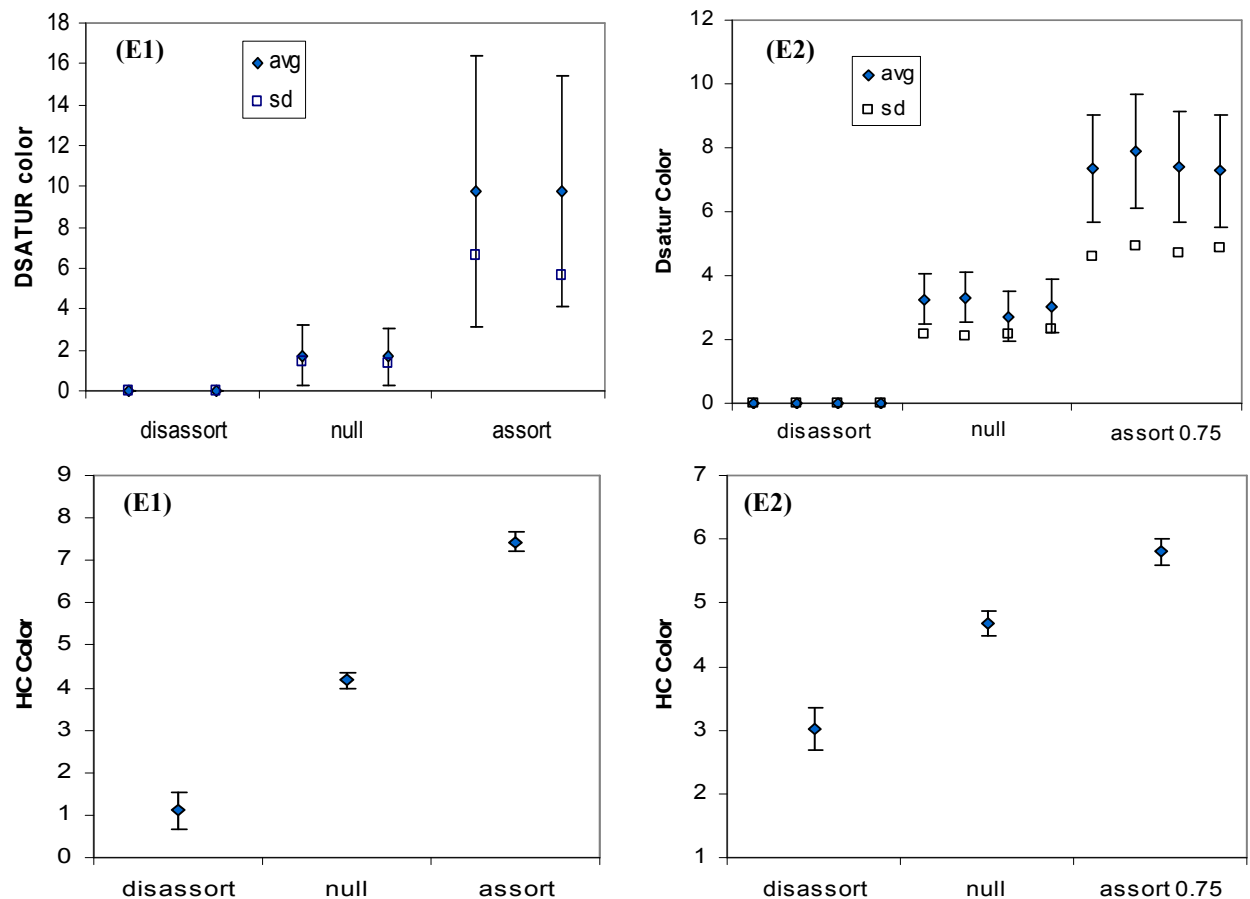

**Fig. SM-1.3** Color summary for top 50 high degree nodes of E1 (left) and E2 (right) networks. Error bars indicate 99% confidence interval. Color range increases significantly as networks become less disassortative (left to right) denoting that more independent sets are created for the same number of nodes.

The plots in Fig. SM-1.3 show that high degree nodes are partitioned into fewer independent sets when a network is less assortative. For both DSATUR and HC, the color range of the top 50 high degree nodes is significantly larger for assortative than disassortative networks. Also, in both E1 and E2 networks, DSATUR colors all the top 50 high degree nodes with the same color 0. This is expected for E2 since link probability is 0.00 between any pair of nodes belonging to the top 50 high degree nodes.

Why are disassortative networks more colorable with a smaller palette? Previously, [17] reported that increases in network clustering increases graph coloring difficulty due to shorter path lengths and increased network cliquishness. Similarly, we find path length amongst nodes of high degree to be a distinguishing factor between disassortative and assortative networks and a determining factor in the number of colors required by DSATUR or by HC. Compared with their assortative counterparts, disassortative networks have longer median path lengths amongst nodes of high degree (q1 MPL) although there is no significant different between median path lengths of the networks as a whole (MPL) (Fig. SM-1.4)

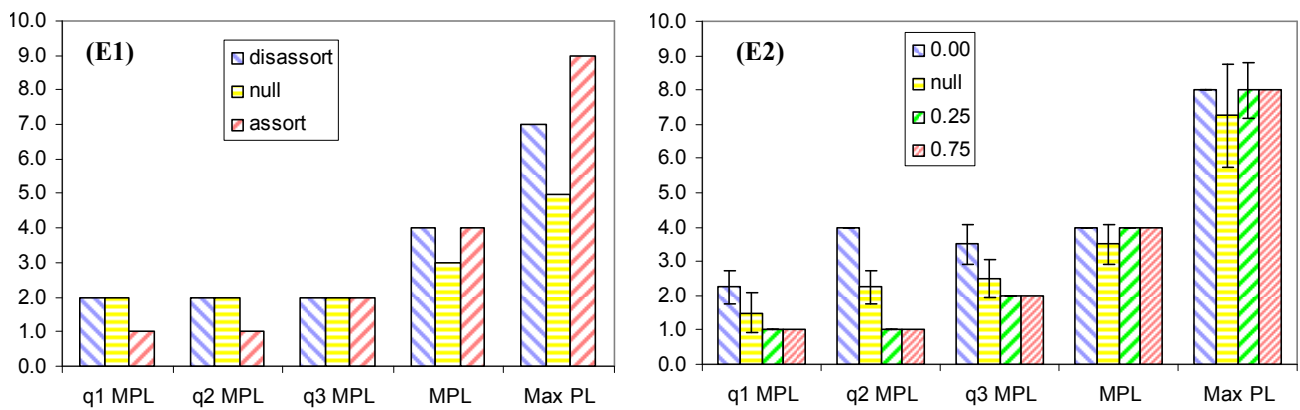

**Fig. SM-1.4** Median path length (MPL) of nodes by degree quartile and average network diameter (Max PL) for E1 networks (left) and for E2 networks (right). Error bars indicate one standard deviation. The quartiles are formed as follows: (i) unique degree values are sorted in ascending order, and (ii) this sorted list is divided into four (almost) equal parts. Quartile 1 (q1) nodes are those with degree values larger than or equal to the minimum value in the upper quartile of this sorted list (Quartile 1 nodes are those with higher degrees). Quartile 2 nodes are those with degree values larger than or equal to the median of this sorted list. Quartile 3 nodes are those with degree values larger than or equal to the minimum value of the lower quartile of this sorted list. Quartile 4 comprises all nodes in the network.

The effect of path length amongst nodes of high degree on graph coloring is intuited as follows: in general, by nature of having more links, nodes with high degree are more constrained in their color choices than nodes with low degree. By preferring to fix the color of high degree nodes, which DSATUR does explicitly in its algorithm and HC does implicitly (negative correlations are recorded between node degree and time of last successful mutation, and between node degree and number of successful mutations), the color palette expands more slowly and less unnecessarily. Nodes of low degree have more color choices and their exact color can be determined later within the existing color range. As such, a network would be colorable with fewer colors if nodes of high degree were separated from each other but still connected to one another via nodes of lower degrees which are less constrained in their color choices. Longer path lengths amongst nodes of high degree reflect networks with such characteristics, as do negative degree-degree correlation or disassortative node degree mixing pattern. Differences in degree-degree correlation may also explain the large performance variation associated with coloring scale-free networks reported in [18].

**SM-2 Supplementary Material for Section 3**

*PPI datafiles*

The PPI networks in this paper are constructed from the data sources listed in Table SM-2.1. These data files are freely available on-line for download and the DIP 2008 dataset was the most recent in the DIP at the time of preparing this paper. Table SM-2.2 lists the organisms in this study. Mammalian does not refer to a particular organism but is included as an additional test network.

Table SM-2.1 PPI data sources

| Label | Details |
|---|---|
| DIP*YYYYMMDD*MIF25 | Species specific Full DIP (http://dip.doe-mbi.ucla.edu) files dated *YYYYMMDD*.MIF25. |
| DIP HiTHr | High throughput datasets in MIF format from DIP (http://dip.doe-mbi.ucla.edu) |
| HPRD | The Human Protein Reference Database (http://www.hprd.org)<br>File used: HPRD_SINGLE_PSIMI_070609.xml |
| TAP | The Yeast TAP Project (http://tap.med.utoronto.ca)<br>Files used: TAP_core.txt and MCL_clusters.txt<br>Krogan et al. Global landscape of protein complexes in the yeast Saccharomyces cerevisiae. *Nature* 2006; 440:637-643. |

Table SM-2.2 Organisms

| Short name | Full name | NCBI TaxId |
|---|---|---|
| Celeg | Caenorhabditis elegans | |
| Dmela | Drosophila melanogaster | |
| Ecoli | Escherichia coli | |
| Hpylo | Helicobacter pylori | |
| Hsapi | Homo sapiens | 9606* |
| Scere | Saccharomyces cerevisiae | |
| Mammalian | Mammalian | 40674* |

* Used to identify interactors and interactions for different organisms in the HPRD file.

Table SM-2.3 gives the size of the PPI datafiles in terms of number of listed interactors and interactions. Self interactions are those with only one distinct interactor listed per interaction. Binary interactions are those with exactly two distinct interactors listed per interaction. Complex interactions are those with more than two distinct interactors listed per interaction.

**Table SM-2.3** Characteristics of PPI data files

| Data source | | Organism | DID | Interactors | Interactions | | |
|---|---|---|---|---|---|---|---|
| | | | | | Binary | Complex | Self |
| DIP HiTHr | Gavin2002a | Scere | 1S | 1,361 | 3,221 | 0 | 0 |
| | Giot2003a | Dmela | 1D | 7,036 | 20,732 | 0 | 193 |
| | Li2004a | Celeg | 1C | 2,633 | 3,966 | 0 | 60 |
| TAP (2006) | | Scere | 3S | 2,708 | 7,123 | 339 | 0 |
| DIP20070219MIF25 | | Celeg | 4C | 2,646 | 3,976 | 0 | 60 |
| | | Dmela | 4D | 7,461 | 22,641 | 1 | 185 |
| | | Ecoli | 4E | 1,858 | 5,928 | 445 | 1,041 |
| | | Hpylo | 4P | 710 | 1,358 | 0 | 61 |
| | | Hsapi | 4H | 1,186 | 1,427 | 13 | 64 |
| | | Scere | 4S | 4,968 | 17,240 | 779 | 289 |
| DIP20081014MIF25 | | Celeg | 5C | 2,651 | 3,979 | 0 | 61 |
| | | Dmela | 5D | 7,505 | 22,677 | 9 | 186 |
| | | Ecoli | 5E | 1,879 | 5,937 | 445 | 1,052 |
| | | Hpylo | 5P | 713 | 1,360 | 0 | 61 |
| | | Hsapi | 5H | 1,645 | 1,806 | 79 | 138 |
| | | Scere | 5S | 4,977 | 17,226 | 801 | 294 |
| HPRD (Release 8, 2009) | | Hsapi | 7H | 3,214 | 3,555 | 9 | 509 |
| | | Mammalian | 7X | 6,148 | 18,523 | 456 | 1,583 |

*PPI network construction*

Interactors and non-self interactions in a PPI datafile become respectively the nodes and links of a PPI network. Except for the TAP dataset, the topology of complex interactions is unspecified in the PPI datafiles. As such, we first use a spanning tree (built by adding one node at a time to the existing tree) to link all nodes participating in a complex interaction, and then use a parameter *Pe* which we introduce to specify the probability of adding links to the complex. Links built in this manner are hypothetical and may coincide with actual interactions or not. The spoke model is another way to handle the undetermined topological aspect of complex interactions but this requires knowledge or selection of a central node (the bait) from which links are made to all other participants of a complex [1]. The choice of *Pe* affects the number of links in a PPI network with complexes, and may also affect node degree and other network statistics such as clustering coefficient, assortativity and path length. As such, three *Pe* values are used in our experiments: 0.00, 0.25 and 0.50.

Interactions in the TAP datafile (TAP_core.txt) are all binary. The 339 complex interactions for TAP are derived from the accompanying MCL_cluster.txt file as follows: for each cluster in MCL_cluster.txt (there are 547 clusters, some with only two members or interactors), retain only interactors found in TAP_core.txt and then count as a complex, only those clusters with more than two members.

Table SM-2.4 summarizes the fixed (*Pe* independent) characteristics of PPI networks generated from the PPI datafiles in Table SM-2.3. The number of nodes in Table SM-2.4 may differ from the number of interactors in Table SM-2.3 because we only include in our PPI networks those interactors listed as participants in an interaction. A complex node is a node participating in a complex interaction or equivalently belonging to a complex. Complex size refers to the number of nodes in a complex. Dividing the number of complex nodes by the number of complexes need not yield average complex size because complexes may overlap, i.e. a complex node may belong to more than one complex, and average complex size counts a shared complex node multiple times.

Table SM-2.5 gives a sample of values for the variable (*Pe* dependent) characteristics of PPI networks. The values may vary only for PPI networks with unspecified topology for complexes (these networks are highlighted in gray).

*Dealing with inaccuracies in PPI data*

To address the possibility of incompleteness and expected high false positive rate in PPI data, we first use the variation over time in the number of nodes, and number and type of interactions per organism as observed in Tables SM-2.4 and SM-2.5 as a source of noise that is more plausible than simply adding and removing nodes and links at random from a network. Second, links of a network are rewired at random with various proportions *Pr*. First 2% of the links are rewired, then another 2%, and finally 6% to make a total of 10%.

Table SM-2.4 Fixed (*Pe* independent) characteristics of PPI networks

| DID | No. of Nodes (a) | No. of Complex Nodes (b) | Complex nodes % (100b/a) | No. of Complexes | Complex Size ||||
|---|---|---|---|---|---|---|---|---|
| | | | | | Min | Max | Avg | Stdev |
| 1S | 1,361 | 0 | 0.00 | 0 | 0 | 0 | 0 | - |
| 1D | 7,027 | 0 | 0.00 | 0 | 0 | 0 | 0 | - |
| 1C | 2,624 | 0 | 0.00 | 0 | 0 | 0 | 0 | - |
| 3S | 2,708 | 2,554 | 94.31 | 339 | 3 | 49 | 6.4 | 5.9 |
| 4C | 2,637 | 0 | 0.00 | 0 | 0 | 0 | 0 | - |
| 4D | 7,451 | 3 | 0.04 | 1 | 3 | 3 | 3 | - |
| 4E | 1,548 | 1,233 | 79.65 | 445 | 3 | 89 | 13.4 | 12.9 |
| 4P | 701 | 0 | 0.00 | 0 | 0 | 0 | 0 | - |
| 4H | 1,173 | 23 | 1.96 | 13 | 3 | 4 | 3.3 | 0.5 |
| 4S | 4,964 | 1,988 | 40.05 | 779 | 3 | 55 | 9.4 | 8.3 |
| 5C | 2,640 | 0 | 0.00 | 0 | 0 | 0 | 0 | - |
| 5D | 7,495 | 27 | 0.36 | 9 | 3 | 5 | 3.6 | 0.7 |
| 5E | 1,561 | 1,233 | 78.99 | 445 | 3 | 89 | 13.4 | 12.9 |
| 5P | 704 | 0 | 0.00 | 0 | 0 | 0 | 0 | - |
| 5H | 1,595 | 166 | 10.41 | 79 | 3 | 5 | 3.3 | 0.6 |
| 5S | 4,971 | 1,983 | 39.89 | 801 | 3 | 55 | 9.4 | 8.2 |
| 7H | 2,231 | 26 | 1.17 | 9 | 3 | 4 | 3.1 | 0.3 |
| 7X | 5,716 | 956 | 16.72 | 456 | 3 | 12 | 3.9 | 1.5 |

Table SM-2.5 Variable (*Pe* dependent) characteristics of PPI networks

| DID | *Pe* = 0.00 || *Pe* = 0.25 || *Pe* = 0.50 ||
|---|---|---|---|---|---|---|
| | No. of Links | Degree Min, Max, Avg., Stdev. | No. of Links | Degree Min, Max, Avg., Stdev. | No. of Links | Degree Min, Max, Avg., Stdev. |
| 1S | 3,221 | 1, 53, 4.7, 5.9 | | | | |
| 1D | 20,732 | 1, 178, 5.9, 9.4 | | | | |
| 1C | 3,966 | 1, 187, 3.0, 7.2 | | | | |
| 3S | 7,123 | 1, 141, 5.3, 7.5 | | | | |
| 4C | 3,976 | 1, 187, 3.0, 7.2 | | | | |
| 4D | 22,642 | 1, 178, 6.1, 9.8 | 22,642 | 1, 178, 6.1, 9.8 | 22,643 | 1, 178, 6.1, 9.8 |
| 4E | 9,047 | 1, 248, 11.7, 26.1 | 16,176 | 1, 412, 20.8, 44.7 | 21,964 | 1, 523, 28.4, 57.8 |
| 4P | 1,358 | 1, 54, 3.9, 5.4 | | | | |
| 4H | 1,443 | 1, 37, 2.5, 3.0 | 1,443 | 1, 37, 2.5, 3.0 | 1,445 | 1, 37, 2.5, 3.0 |
| 4S | 22,178 | 1, 283, 8.9, 13.8 | 31,862 | 1, 283, 12.8, 20.4 | 40,771 | 1, 321, 16.4, 27.6 |
| 5C | 3,979 | 1, 187, 3.0, 7.2 | | | | |
| 5D | 22,689 | 1, 178, 6.1, 9.8 | 22,693 | 1, 178, 6.1, 9.8 | 22,694 | 1, 178,6.1, 9.8 |
| 5E | 9,087 | 1, 252, 11.6, 26.0 | 16,195 | 1, 428, 20.7, 44.3 | 21,895 | 1, 524, 28.1, 57.4 |
| 5P | 1,360 | 1, 54, 3.9, 5.4 | | | | |
| 5H | 1,892 | 1, 37, 2.4, 2.8 | 1,904 | 1, 37, 2.4, 2.8 | 1,920 | 1, 37, 2.4, 2.8 |
| 5S | 22,158 | 1, 283, 8.9, 13.8 | 31,737 | 1, 283, 12.8, 20.2 | 40,719 | 1, 309, 16.4, 27.5 |
| 7H | 3,561 | 1, 97, 3.2, 6.2 | 3,563 | 1, 97, 3.2, 6.2 | 3,564 | 1, 97, 3.2, 6.2 |
| 7X | 19,181 | 1, 191, 6.7, 10.9 | 19,397 | 1, 190, 6.8, 11.0 | 19,614 | 1, 188, 6.9, 11.0 |

Cells are left blank if there is no change in value.

*PPI network naming convention*

To ease the identification of PPI networks and their variations in the results, we assign numerical labels (NID) to the PPI networks as follows: NID = ODID + $Pe$ + $Pr$. For instance, the NID of a PPI network for *S. cerevisiae* built from dataset DIP20081014MIF25 with $Pe$ = 0.25 and $Pr$ = 0.04 is 4.29. ODID (Table SM-2.6) arranges the networks by data file chronological order and by organism. $Pe$ for networks without complex interactions is 0.00.

**Table SM-2.6** ODID

| ODID | 1 | 2 | 3 | 4 | 5 | 6 | 7 | 8 | 9 | 10 | 11 | 12 | 13 | 14 | 15 | 16 | 17 | 18 |
|---|---|---|---|---|---|---|---|---|---|---|---|---|---|---|---|---|---|---|
| DID | 1S | 3S | 4S | 5S | 1D | 4D | 5D | 1C | 4C | 5C | 4E | 5E | 4P | 5P | 4H | 5H | 7H | 7X |

**SM-3 Supplementary Material for Section 4**

*Network clusters and Protein complexes*

Protein complexes often form network clusters, i.e. densely linked subgraphs, and network clustering forms the basis of protein complex detection algorithms such as HCS [13], MCODE [1] and RNSC [10]. Wherever possible, we use the term 'complex' for a biologically meaningful cluster of protein nodes which has been tagged as such, and 'cluster' for a group of nodes with high link density. A cluster need not be a complex.

Complex interactions are considered as protein complexes. However, this does not mean that there are no protein complexes in PPI networks with no complex interactions specified. The protein complexes in these networks, e.g. 1S and 1D, are just not explicitly identified as such in their datafiles, and we exclude them from our work in section 4 of the paper. Information about protein complexes for 1S and 1D can be derived from other biological databases e.g. MIPS. But we decided to test more recent PPI networks and these have complex interactions explicitly defined in their datafiles. No doubt there are other means of creating PPI networks and discovering their complexes, e.g. combining different data sources, but these are not dealt with in our current work.

*Results*

Fig. SM-3.1 compares pairs of corresponding after the results are summarized by ODID and *Pe*.

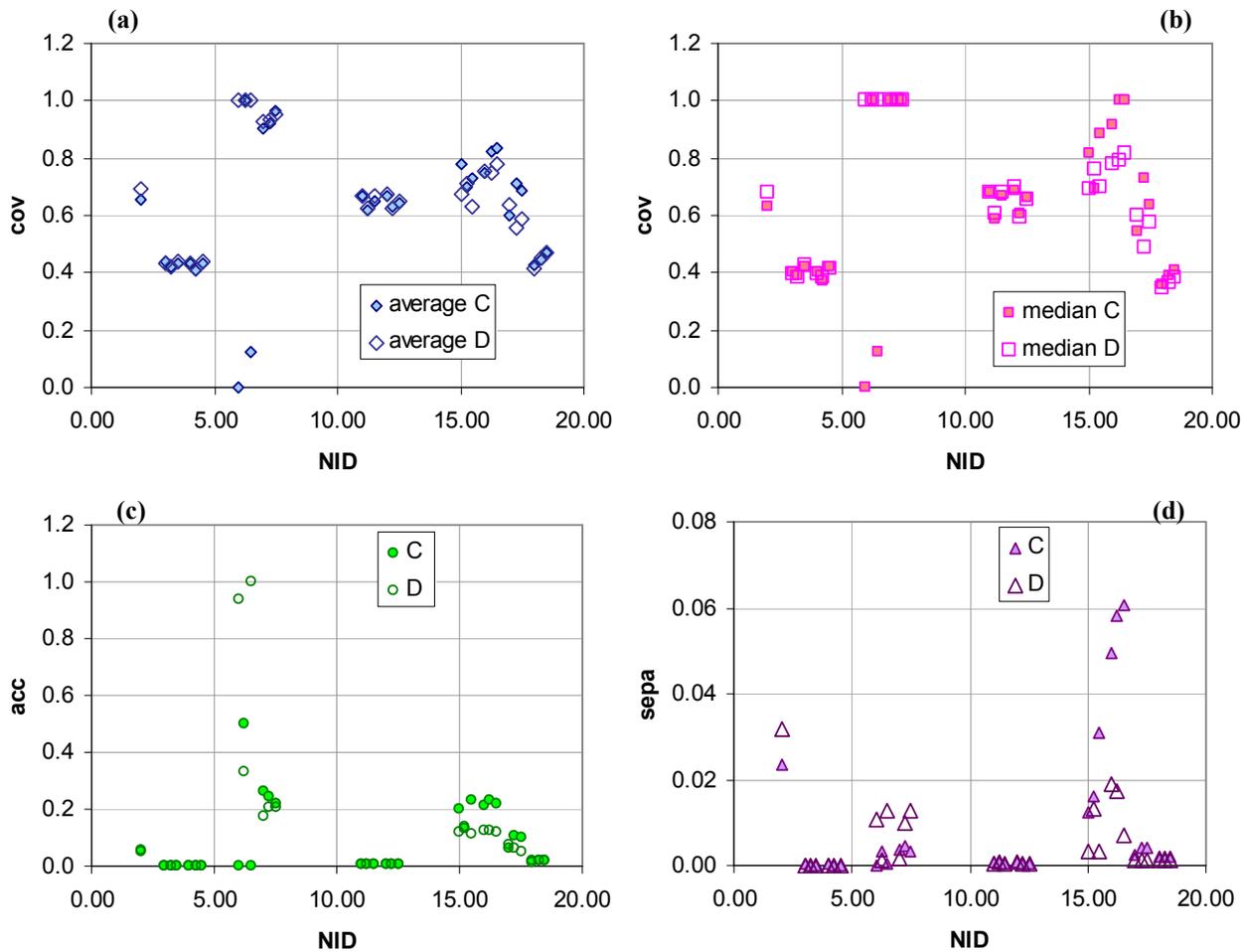

**Fig. SM-3.1** Comparison of the 'after' clustering quality statistics

Tables SM-3.1a and SM-3.1b contain data produced for 5S (*S. cerevisiae* from DIP20081014MIF25) to illustrate the summarization process described in section 4.3. The ODID for 5S is 4. **Col** notes the average values (Avg.) compared in Figs. 9 and SM-3.1.

Table SM-3.1a Basic algorithm (Avg. Color, 'C') results for 5S

| NID | Before | | | | After | | | | After - Before | | | | ODID + *Pe* |
|---|---|---|---|---|---|---|---|---|---|---|---|---|---|
| | Avg. cov | Median cov | Acc | Sepa | Avg. cov | Median cov | Acc | Sepa | Avg. cov | Median cov | Acc | Sepa | |
| Col | | | | | SM-3.1a | SM-3.1b | SM-3.1c | SM-3.1d | 9b | 9c | 9d | 9e | |
| 4.00 | 0.4662 | 0.4366 | 0.0026 | 0.0002 | 0.4769 | 0.4463 | 0.0028 | 0.0003 | 0.0107 | 0.0097 | 0.0002 | 0.0001 | |
| 4.02 | 0.4401 | 0.4150 | 0.0024 | 0.0002 | 0.4532 | 0.4281 | 0.0029 | 0.0003 | 0.0131 | 0.0131 | 0.0005 | 0.0001 | |
| 4.04 | 0.4182 | 0.3928 | 0.0021 | 0.0002 | 0.4301 | 0.3995 | 0.0027 | 0.0003 | 0.0119 | 0.0067 | 0.0006 | 0.0001 | |
| 4.10 | 0.3526 | 0.3212 | 0.0015 | 0.0001 | 0.3628 | 0.3320 | 0.0016 | 0.0003 | 0.0102 | 0.0108 | 0.0001 | 0.0002 | |
| Avg. | 0.4193 | 0.3914 | 0.0022 | 0.0002 | **0.4308** | **0.4015** | **0.0025** | **0.0003** | **0.0115** | **0.0101** | **0.0004** | **0.0001** | 4.00 |
| 4.25 | 0.4428 | 0.4090 | 0.0022 | 0.0002 | 0.4502 | 0.4154 | 0.0023 | 0.0002 | 0.0074 | 0.0064 | 0.0001 | 0.0000 | |
| 4.27 | 0.4182 | 0.3796 | 0.0016 | 0.0001 | 0.4279 | 0.3890 | 0.0018 | 0.0002 | 0.0097 | 0.0094 | 0.0002 | 0.0001 | |
| 4.29 | 0.4000 | 0.3654 | 0.0016 | 0.0001 | 0.4079 | 0.3659 | 0.0017 | 0.0002 | 0.0079 | 0.0005 | 0.0001 | 0.0001 | |
| 4.35 | 0.3432 | 0.3111 | 0.0011 | 0.0001 | 0.3534 | 0.3261 | 0.0012 | 0.0001 | 0.0102 | 0.0150 | 0.0001 | 0.0000 | |
| Avg. | 0.4011 | 0.3663 | 0.0016 | 0.0001 | **0.4099** | **0.3741** | **0.0018** | **0.0002** | **0.0088** | **0.0078** | **0.0001** | **0.0001** | 4.25 |
| 4.50 | 0.4635 | 0.4602 | 0.0023 | 0.0001 | 0.4701 | 0.4642 | 0.0023 | 0.0002 | 0.0066 | 0.0040 | 0.0000 | 0.0001 | |
| 4.52 | 0.4381 | 0.4296 | 0.0019 | 0.0001 | 0.4448 | 0.4341 | 0.0020 | 0.0002 | 0.0067 | 0.0045 | 0.0001 | 0.0001 | |
| 4.54 | 0.4223 | 0.4116 | 0.0018 | 0.0001 | 0.4296 | 0.4248 | 0.0019 | 0.0002 | 0.0073 | 0.0132 | 0.0001 | 0.0001 | |
| 4.60 | 0.3662 | 0.3432 | 0.0011 | 0.0001 | 0.3752 | 0.3512 | 0.0014 | 0.0001 | 0.0090 | 0.0080 | 0.0003 | 0.0000 | |
| Avg. | 0.4225 | 0.4112 | 0.0018 | 0.0001 | **0.4299** | **0.4186** | **0.0019** | **0.0002** | **0.0074** | **0.0074** | **0.0001** | **0.0001** | 4.50 |

Table SM-3.1b Alternative algorithm (Degree, 'D') results for 5S

| NID | Before | | | | After | | | | After - Before | | | | ODID + *Pe* |
|---|---|---|---|---|---|---|---|---|---|---|---|---|---|
| | Avg. cov | Median cov | Acc | Sepa | Avg. cov | Median cov | Acc | Sepa | Avg. cov | Median cov | Acc | Sepa | |
| Col | | | | | SM-3.1a | SM-3.1b | SM-3.1c | SM-3.1d | 9b | 9c | 9d | 9e | |
| 4.00 | 0.4665 | 0.4425 | 0.0025 | 0.0002 | 0.4765 | 0.4459 | 0.0026 | 0.0002 | 0.0100 | 0.0034 | 0.0001 | 0.0000 | |
| 4.02 | 0.4411 | 0.4099 | 0.0022 | 0.0002 | 0.4527 | 0.4175 | 0.0023 | 0.0002 | 0.0116 | 0.0076 | 0.0001 | 0.0000 | |
| 4.04 | 0.4162 | 0.3793 | 0.0020 | 0.0002 | 0.4295 | 0.3884 | 0.0021 | 0.0002 | 0.0133 | 0.0091 | 0.0001 | 0.0000 | |
| 4.10 | 0.3518 | 0.3111 | 0.0014 | 0.0001 | 0.3679 | 0.3333 | 0.0015 | 0.0002 | 0.0161 | 0.0222 | 0.0001 | 0.0001 | |
| Avg. | 0.4189 | 0.3857 | 0.0020 | 0.0002 | **0.4317** | **0.3963** | **0.0021** | **0.0002** | **0.0128** | **0.0106** | **0.0001** | **0.0000** | 4.00 |
| 4.25 | 0.4473 | 0.4246 | 0.0022 | 0.0002 | 0.4578 | 0.4286 | 0.0022 | 0.0002 | 0.0105 | 0.0040 | 0.0000 | 0.0000 | |
| 4.27 | 0.4210 | 0.3788 | 0.0019 | 0.0002 | 0.4324 | 0.3920 | 0.0020 | 0.0002 | 0.0114 | 0.0132 | 0.0001 | 0.0000 | |
| 4.29 | 0.4003 | 0.3636 | 0.0017 | 0.0001 | 0.4139 | 0.3694 | 0.0018 | 0.0002 | 0.0136 | 0.0058 | 0.0001 | 0.0001 | |
| 4.35 | 0.3468 | 0.3041 | 0.0012 | 0.0001 | 0.3648 | 0.3333 | 0.0013 | 0.0001 | 0.0180 | 0.0292 | 0.0001 | 0.0000 | |
| Avg. | 0.4039 | 0.3678 | 0.0018 | 0.0002 | **0.4172** | **0.3808** | **0.0018** | **0.0002** | **0.0134** | **0.0131** | **0.0001** | **0.0000** | 4.25 |
| 4.50 | 0.4687 | 0.4563 | 0.0020 | 0.0002 | 0.4797 | 0.4712 | 0.0021 | 0.0002 | 0.0110 | 0.0149 | 0.0001 | 0.0000 | |
| 4.52 | 0.4410 | 0.4333 | 0.0018 | 0.0001 | 0.4542 | 0.4443 | 0.0018 | 0.0002 | 0.0132 | 0.0110 | 0.0000 | 0.0001 | |
| 4.54 | 0.4227 | 0.4070 | 0.0018 | 0.0001 | 0.4359 | 0.4144 | 0.0018 | 0.0002 | 0.0132 | 0.0074 | 0.0000 | 0.0001 | |
| 4.60 | 0.3591 | 0.3347 | 0.0012 | 0.0001 | 0.3750 | 0.3399 | 0.0012 | 0.0001 | 0.0159 | 0.0052 | 0.0000 | 0.0000 | |
| Avg. | 0.4229 | 0.4078 | 0.0017 | 0.0001 | **0.4362** | **0.4175** | **0.0017** | **0.0002** | **0.0133** | **0.0096** | **0.0000** | **0.0001** | 4.50 |